\documentstyle[preprint,aps,epsf,floats]{revtex} 

\begin{document}

\preprint{\tighten \vbox{\hbox{FERMILAB-Pub-99/025-T} \hbox{UTPT-99-03} 
  \hbox{UCSD/PTH 99-03} \hbox{CALT-68-2211} \hbox{hep-ph/9903305} }}

\title{The $\bar B\to X_s \gamma$ Photon Spectrum}

\author{Zoltan Ligeti,${}^{a}$ Michael Luke,${}^{b}$
  Aneesh V.\ Manohar,${}^c$ and Mark B.\ Wise${\,}^d$ }

\address{ 
  $^a$Theory Group, Fermilab, P.O.\ Box 500, Batavia, IL 60510 \\[-4pt]
  $^b$Department of Physics, University of Toronto, \\[-8pt]
    60 St. George Street, Toronto, Ontario, Canada M5S 1A7 \\[-4pt]
  $^c$Department of Physics, University of California at San Diego, \\[-8pt]
    9500 Gilman Drive, La Jolla, CA 92093--0319 \\[-4pt]
  $^d$California Institute of Technology, Pasadena, CA 91125  } 

\maketitle

{\tighten
\begin{abstract}%
The photon energy spectrum in inclusive weak radiative $\bar B\to X_s\gamma$
decay is computed to order $\alpha_s^2 \beta_0$.  This result is used to
extract a value for the HQET parameter $\bar\Lambda$ from the average $\langle
1-2E_\gamma / m_B \rangle$, and a value of the parameter $\lambda_1$ from
$\langle (1-2E_\gamma / m_B)^2 \rangle$.  An accurate measurement of $\langle
1-2E_\gamma / m_B \rangle$ can determine the size of the nonperturbative
contributions to the $\Upsilon(1S)$ mass which cannot be absorbed into the $b$
quark pole mass.

\end{abstract}
}

\newpage

Comparison of the measured weak radiative $\bar B\to X_s\gamma$ decay rate with
theory is an important test of the standard model.  In contrast to the decay
rate itself, the shape of the photon spectrum is not expected to be sensitive
to new physics, but it can nevertheless provide important information.  First
of all, studying the photon spectrum is important for understanding how
precisely the total rate can be predicted in the presence of an experimental
cut on the photon energy~\cite{CLEO}, which is important for a model
independent interpretation of the resulting decay rate.  Secondly, moments of
the photon spectrum may be used to measure the heavy quark effective theory
(HQET) parameters which determine the quark pole mass and kinetic
energy~\cite{FLS,AZ}, much like the shape of the lepton energy~\cite{gremmetal}
or hadronic invariant mass~\cite{FLSmass} spectrum in semileptonic $\bar B\to
X_c\,\ell\,\bar\nu$ decay.  The main purpose of this paper is to present the
order $\alpha_s^2\beta_0$ piece of the two-loop correction to the photon
spectrum, and to study its implications.  A calculation to this order is
required for a meaningful comparison of the HQET parameters extracted from
$\bar B\to X_s\gamma$ with those from other processes.

To leading order in small weak mixing angles the effective Hamiltonian is
\begin{equation}
H_{\rm eff} = - \frac{4G_F}{\sqrt2}\, V_{tb} V_{ts}^* 
  \sum_{i=1}^8 C_i(\mu)\, O_i \,,
\end{equation}
where $G_F$ is the Fermi constant, $V_{ij}$ are elements of the
Cabibbo--Kobayashi--Maskawa matrix, $C_i (\mu)$ are Wilson coefficients
evaluated at a subtraction point $\mu$, and $O_i$ are the dimension six
operators
\begin{equation}\label{ops}
\begin{array}{rclrcl}
O_1 &=& (\bar c_{L\beta} \gamma^\mu b_{L\beta}) (\bar s_{L\alpha} \gamma_\mu
  c_{L\beta}) \,, &
O_2 &=& (\bar c_{L\alpha} \gamma^\mu b_{L\alpha}) (\bar s_{L\beta} \gamma_\mu
  c_{L\beta}) \,, \\
O_3 &=& \displaystyle (\bar s_{L\alpha} \gamma^\mu b_{L\alpha}) 
  \sum_q (\bar q_{L\beta} \gamma_\mu q_{L\beta}) \,,\qquad &
O_4 &=& \displaystyle (\bar s_{L\alpha} \gamma^\mu b_{L\beta}) 
  \sum_q (\bar q_{L\beta} \gamma_\mu q_{L\alpha}) \,, \\
O_5 &=& \displaystyle (\bar s_{L\alpha} \gamma^\mu b_{L\alpha}) 
  \sum_q (\bar q_{R\beta} \gamma_\mu q_{R\beta}) \,, &
O_6 &=& \displaystyle (\bar s_{L\alpha} \gamma^\mu b_{L\beta}) 
  \sum_q (\bar q_{R\beta} \gamma_\mu q_{R\alpha}) \,, \\
O_7 &=& \displaystyle {e\over 16\pi^2}\, 
  m_b \bar s_{L\alpha} \sigma^{\mu\nu} b_{R\alpha} F_{\mu\nu} \,, &
O_8 &=& \displaystyle {g\over16\pi^2}\,
  m_b \bar s_{L\alpha} \sigma^{\mu\nu}
  T_{\alpha\beta}^a b_{R\beta} G_{\mu\nu}^a \,.
\end{array}
\end{equation}
In Eq.~(\ref{ops}), $e$ is the electromagnetic coupling, $g$ is the strong
coupling, $m_b$ is the $b$ quark mass, $F_{\mu\nu}$ is the electromagnetic
field strength tensor, $G_{\mu\nu}^a$ is the strong interaction field strength
tensor, and $T^a$ is a color $SU(3)$ generator.  The sums over $q$ include $q =
u,d,s,c,b$ and the subscripts $L,R$ denote left and right handed fields.  The
Wilson coefficients have been calculated to next-to-leading order
(NLO)~\cite{Misiak,match,fourquark}.  Using $\alpha_s(m_Z)=0.12$, and the
convention that the covariant derivative is $D_\mu = \partial_\mu +
igA_\mu^aT^a + ie\,QA_\mu$ (where $Q$ is the fermion's electric charge), the
values we need are  $C_2(m_b)=1.13$, $C_7(m_b)=-0.306$,
$C_8(m_b)=-0.168$~\cite{Misiak}.  

For the photon energy, $E_\gamma$, not too close to its maximal value, the
photon spectrum $d\Gamma/dE_\gamma$ for weak radiative $B$ decay has a
perturbative expansion in the strong interaction fine structure constant
$\alpha_s$.  It is known at order $\alpha_s$ and the main purpose of this
letter is to present the order $\alpha_s^2 \beta_0$ (so-called BLM~\cite{BLM})
contribution.  It is well known that the part of the order $\alpha_s^2$ piece
proportional to the one-loop beta function, $\beta_0 = 11-2n_f/3$ usually
provides a reliable estimate of the full order $\alpha_s^2$ piece.  This part
of the order $\alpha_s^2$ contribution is straightforward to compute using the
method of Smith and Voloshin~\cite{SmVo}.

Using the dimensionless variable\footnote{Later we will introduce a
dimensionless photon energy variable normalized by the $B$ meson mass, $x_B = 2
E_\gamma/m_B$.}, $x_b = 2E_\gamma/m_b$, the photon energy spectrum in $\bar
B\to X_s \gamma$ takes the form
\begin{equation}\label{spectrum}
\frac1{\Gamma_0}\, \frac{d\Gamma}{dx_b} \bigg|_{x_{b} < 1} 
  = A_0 (x_b) + \frac{\alpha_s (m_b)}{\pi} A_1 (x_b) 
  + \bigg( \frac{\alpha_s (m_b)}{\pi} \bigg)^2 \beta_0\, A_2(x_b) + \ldots \,,
\end{equation}
where 
\begin{equation}
\Gamma_0 = {G_F^2\,|V_{tb}V_{ts}^*|^2\,\alpha_{\rm em}\,C_7^2\over32\pi^4}\, 
  m_b^5 \,,
\end{equation}
is the contribution of the tree level matrix element of $O_7$ to the 
$B\to X_s\gamma$ decay rate, and
\begin{equation}\label{Apdef}
A_p(x_b) = \sum_{i\leq j} a_p^{ij} (x_b) 
  \Bigg[ \frac{C_i (m_b) C_j (m_b)}{C_7 (m_b)^2} \Bigg] \,.
\end{equation}
The sums over $i,j$ in Eq.~(\ref{Apdef}) give the contributions of the various
operators in Eq.~(\ref{ops}) to the photon energy spectrum.

It is important to note that since the coefficients in $H_{\rm eff}$ are known
only to NLO accuracy, the BLM calculation of the $O_1-O_8$ contribution to the
photon spectrum is only meaningful away from the endpoint.  At the endpoint,
order $\alpha_s^2$ contributions to the matrix elements are the same order as
the unknown NNLO running [where $\alpha_s\ln(m_W/m_b)$ is counted as ${\cal
O}(1)$].  Neglecting the small contribution to $A_0$ from $O_1-O_6$ discussed
in the next paragraph, at least one gluon must be in the final state to
populate the spectrum for $x_b<1$, so it is consistent to combine the
$\alpha_s^2$ matrix elements with the NLO Wilson coefficients.  (Strictly
speaking, we should for consistency only use the $\beta_0$ part of the NLO
running of the operators with the BLM calculation, but for simplicity we will
use the full NLO result.  The difference between these two approaches is
small.)  Thus powers of $\alpha_s$ in Eq.~(\ref{spectrum}) and elsewhere
reflect the perturbation expansion of the matrix elements only, and not of the
Wilson coefficients.

At zeroth order in the strong coupling, the spectrum for $x_b<1$ arises from
matrix elements of the four-quark operators $O_1-O_6$ in Eq.~(\ref{ops}).  Of
these $O_1$ and $O_2$ include two charm quarks in the final state, and
therefore they contribute to the photon spectrum only for lower values of $x_b$
than what we consider in this paper.  These contributions are divergent in
perturbation theory, and the divergence can be absorbed into the definition of
the quark to photon fragmentation function, $D^{q\to\gamma}(x)$, which depends
on an infrared scale $\Lambda$.  $D^{q\to\gamma}(x)$ is calculable in the
leading logarithmic approximation~\cite{EW,decayfn}.  There is some data on
$D^{q\to \gamma}(x)$, however, the experimental errors are still quite
large~\cite{ALEPH}.  This fragmentation contribution to the coefficients
$a_0^{ij} (x)$ vanishes as $x_b\to 1$, and it is small in the region of large
$x_b$, $0.65<x_b$, which we consider in this paper.  

\begin{figure}
\centerline{\epsfysize=7cm\epsffile{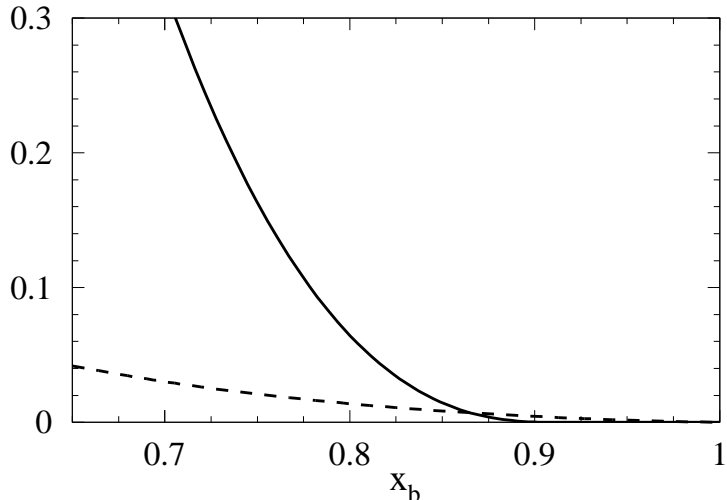}}
\tighten{\caption{$B$ decay background to the photon spectrum due to the 
operators $(\bar c_L \gamma^\mu b_L) (\bar d_L \gamma_\mu u_L)$ (solid curve) 
and $(\bar u_L \gamma^\mu b_L) (\bar d_L \gamma_\mu u_L)$ (dashed curve). }}
\end{figure}

A very important $B$ decay background to the $\bar B\to X_s\gamma$ photon
spectrum is from nonleptonic $b\to c\bar ud$ and $b\to u\bar ud$ decays, where
a massless quark in the final state radiates a photon.  Such backgrounds due to
the operators $(\bar c_L \gamma^\mu b_L) (\bar d_L \gamma_\mu u_L)$ and $(\bar
u_L \gamma^\mu b_L) (\bar d_L \gamma_\mu u_L)$ are shown in Fig.~1 (using
$|V_{ub}/V_{cb}|=0.1$).  We used the Duke--Owens parameterization of the
fragmentation function~\cite{DO}, setting $\Lambda=1.3\,$GeV and $Q^2=m_b^2$. 
(This value of $\Lambda$ is motivated by a fit to the ALEPH data~\cite{ALEPH}.)
The uncertainty of this result is sizable, since the $\Lambda$-dependence is
large and $m_b$ may not be large enough to justify keeping only the leading
logarithms.  Close to maximal $x_b$ the resummed fragmentation function may
predict too large a suppression of the photon spectrum, since the lightest
exclusive final states dominate there.  The background from $b\to c \bar u d$
($b\to u \bar u d$) is more than 50\% of the 77 contribution to
$(1/\Gamma_0)d\Gamma/dx_b$ below $x_b\sim0.75$ ($x_b\sim0.65$).\footnote{Note
that these backgrounds are steeply falling functions of $x_b$, and are indeed
negligible in the present CLEO region of $E_\gamma>2.1\,$GeV.  The tree level
contribution of the operators $O_3-O_6$ in Eq.~(\ref{ops}) to the photon
spectrum is about a fifth of the $b\to u \bar u d$ background.}  Therefore, we
will concentrate on the region $x_b>0.65$; to measure the $B\to X_s\gamma$
photon spectrum at lower values of $x_b$ would not only require excluding final
states with charm with very good efficiency, but also demanding a strange quark
in the final state.  Note that for $B\to X_d\gamma$, the fragmentation
contribution from $b\to u\bar ud$ is larger than the short distance piece
unless $x_b$ is very close to~1.

Neglecting the strange quark mass, $a_1^{88}$ is also divergent in perturbation
theory.  This divergence can also be absorbed into the definition of
fragmentation functions.  In the leading logarithmic approximation~\cite{KLP}
\begin{equation}
a_1^{88} (x) = \left(\frac{4\pi}{3\alpha_{\rm em}}\right) 
  [ D^{s\to \gamma}(x) + D^{g\to \gamma}(x) ] \,,
\end{equation}
where $D^{s\to\gamma}(x)$ and $D^{g\to\gamma}(x)$ are the strange quark to
photon and gluon to photon fragmentation functions, which have large
uncertainties.  In the region $x_b>0.65$, the $a_1^{88}$ contribution to the
photon spectrum $(1/\Gamma_0)d\Gamma/dx_b$ is less than 0.01.  Given the
uncertainty in $a_1^{88}$, and its small magnitude, it does not appear useful
to calculate $a_2^{88}$.

Experimentally, because of backgrounds, only $\bar B\to X_s\gamma$ photons with
large energies can be detected.  The present experimental cut is $E_\gamma >
2.1\,$GeV at CLEO~\cite{CLEO}, which corresponds to $x_b>0.875$ with
$m_b=4.8\,$GeV.  In the large $x_b$ region the most important contribution to
the sum in Eq.~(\ref{Apdef}) come from the 77 term, with moderate corrections
from the 22, 78, and 27 terms.  The other contributions (88, 28, and the ones
involving $O_1$ and $O_3 - O_6$) are very small, and will be neglected in this
paper.

Simple analytic expressions for $a_1^{77}$ and $a_1^{78}$ are 
available,
\begin{eqnarray}
a_1^{77} (x) &=& \frac{(2x^2 - 3x - 6) x + 2(x^2 - 3) \ln(1-x)}{3(1-x)} \,,
  \label{a177} \\
a_1^{78} (x) &=& \frac{8}{9} \bigg[ \frac{4 + x^2}{4} + \frac{1-x}{x} 
  \ln(1-x) \bigg] \,. \label{a178}
\end{eqnarray}
Neglecting the small $A_0$ term in Eq.~(\ref{spectrum}), we can calculate the
shape of the photon spectrum away from $x=1$ to order $\alpha_s^2\beta_0$
accuracy knowing the effective Hamiltonian to order $\alpha_s$ 
(NLO) only.  At order $\alpha_s^2\beta_0$, we find that 
$a_2^{77}$ and $a_2^{78}$ are given by
\begin{eqnarray}
a_2^{77}(x) &=& \frac{1}{18} \Bigg[ \frac{38x^3 - 93x^2 + 6x - 36}{4(1-x)} 
  - \frac{6x^4 - 31 x^3 + 24x^2 - 30x + 18}{2x (1-x)} \ln (1-x) \nonumber\\*
&& + 3 (3 - x^2) \frac{3\ln^2 (1 - x) + 2L_2 (x)}{2 (1 - x)}\Bigg] \,, 
  \label{a277} \\
a_2^{78}(x) &=& \frac{1}{9} \Bigg[ \frac{19x^2 - 24x + 88}{12} 
  - \frac{3x^3 - 12x^2 + 56x - 32}{6x} \ln(1-x) \nonumber\\*
&& - (1 - x) \frac{3\ln^2 (1 - x) + 2L_2 (x)}{x} \Bigg] \,, \label{a278}
\end{eqnarray}
where $L_2(z) = -\int_0^z dt \ln(1-t)/t$ is the dilogarithm.  The strange quark
mass is neglected throughout this paper; it only enters the final results
quadratically, as \mbox{$m_s^2 / [m_b^2(1-x_b)]$}.

The functions of $a_1^{22}$ and $a_1^{27}$ are known in the
literature~\cite{AG,Pott}, and we agree analytically with those results.  The
order $\alpha_s^2\beta_0$ contributions, $a_2^{22}$ and $a_2^{27}$, are
computed numerically.  We find it most useful to present simple approximations
to these functions
\begin{eqnarray}\label{ai22}
a_1^{22}(x) &\simeq& -0.0842 + 0.3333x - 0.2005x^2 + 0.0227x^3 \nonumber\\*
&&\phantom{} + \bigg({m_c\over m_b} - {1.4\over4.8}\bigg)
  (-0.454 + 0.061x) \,, \nonumber\\*
a_2^{22}(x) &\simeq& -0.1272 + 0.3957x - 0.3227x^2 + 0.0952x^3
  - 0.0180\ln(1-x) \nonumber\\*
&&\phantom{} + \bigg({m_c\over m_b} - {1.4\over4.8}\bigg)
  [-0.155 - 0.106x + 0.106\ln(1-x)] \,, 
\end{eqnarray}
and
\begin{eqnarray}\label{ai27}
a_1^{27}(x) &\simeq& -0.1064 + 0.4950x - 0.4361x^2 + 0.0373x^3  \nonumber\\*
&&\phantom{} + \bigg({m_c\over m_b} - {1.4\over4.8}\bigg)
  (-1.207 + 2.901x) \,, \nonumber\\*
a_2^{27}(x) &\simeq& -0.0156 + 0.0463x + 0.3467x^2 - 0.3045x^3 
  + 0.0027\ln(1-x) \nonumber\\*
&&\phantom{} + \bigg({m_c\over m_b} - {1.4\over4.8}\bigg)
  [-1.523 + 2.538x - 0.448\ln(1-x)] \,.
\end{eqnarray}
These approximations are accurate to within 1\% in the region $x_b>0.6$ for
$m_c/m_b = 1.4/4.8$.  The 27 contribution is very sensitive to $m_c/m_b$. 
Changing $m_c/m_b$ from $1.4/4.8$ to $1.2/4.6$ or $1.6/5.0$ modifies $a_1^{27}$
and $a_2^{27}$ dramatically.  The 22 contribution only changes in the
previously mentioned range of $m_c/m_b$ by $\pm(20-25)\%$.  The 22 contribution
is also accurate to within 1\% when $m_c/m_b$ changes by $\pm0.03$. 
However, the 27 contribution is only accurate at the 20\% level when $m_c/m_b$
changes in this range.  Note that the perturbation series in $\alpha_s$ is
particularly badly behaved for the 27 contribution.  Roughly 2/3 of the 22
contribution is from absorptive parts corresponding to real intermediate states.

The coefficients $a_2^{ij}$ are determined by calculating the order $\alpha_s^2
n_f$ piece and making the identification, $-2n_f/3 \to \beta_0$.  There is a
subtlety in applying this method to weak radiative $B$ decay.  There is a
contribution of order $\alpha_s^0 n_f$ from the tree level $b \to s\gamma q
\bar q$ matrix elements of $O_3 - O_6$, coming from Feynman diagrams where the
photon couples to the bottom or strange quarks.  It is not associated with a
term of order $\alpha_s^0 \beta_0$.  To avoid adding an analogous spurious
order $\alpha_s^2 \beta_0$ contribution to $a_2^{27}$ and $a_2^{22}$, only
diagrams where the photon couples to the charm quark were included in the
calculation of the matrix element of $O_2$.

Part of the $\bar B\to X_s\gamma$ matrix element of $O_2$ is not adequately
calculated in perturbation theory.  It corresponds to the process $\bar B\to
J/\psi X_s$ followed by the decay $J/\psi\to \gamma + {\rm (light\ hadrons)}$. 
There will be large corrections to the part of the charm quark loop where the
$c\bar c$ are almost on-shell and have the same velocity.  In this region there
are large ``Coulombic QCD corrections'' that produce the $J/\psi$ state. 
However, cutting this small part of the $c\bar c$ phase space out of our
calculation of the matrix element of $O_2$ has a negligible effect.  Hence, at
the order of perturbation theory to which we are working, calculating the
$c\bar c$ loop while removing $J/\psi$'s from the data would be a consistent
approximation.

The sum of the 77, 22, 78, and 27 contributions is plotted in Fig.~2 in the
region $0.65<x_b<0.9$ (using $\alpha_s(m_b)=0.22$ and $\beta_0=25/3$).  For
very large $x$, other effects that we have not calculated become important. 
There are both nonperturbative and perturbative terms that are singular as
$x\to 1$.  They sum into a shape function that modifies the spectrum in this
region~\cite{shapefn}.  Unfortunately, at the present time, it is not possible
to make a model independent estimate of these effects.  Therefore, we do not
plot the perturbation theory predictions for $x_b>0.9$.  In the plotted region,
the 22, 78, and 27 terms make a moderate correction to the dominant 77
contribution to $(1/\Gamma_0)d\Gamma/dx$, which is shown in Fig.~2 with the
thin curves.

\begin{figure}
\centerline{\epsfysize=7cm\epsffile{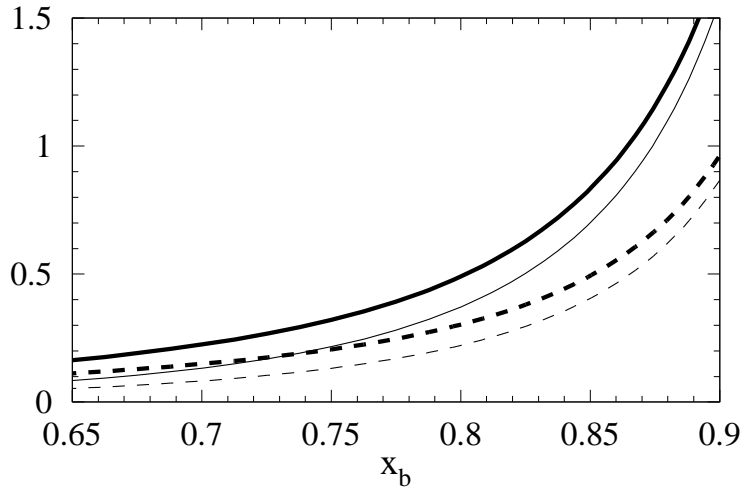}}
\tighten{\caption{The sum of the 77, 22, 78, and 27 contributions to 
$(1/\Gamma_0)d\Gamma/dx_b$ at order $\alpha_s$ (thick dashed curve) and 
$\alpha_s^2\beta_0$ (thick solid curve).  The thin curves show the 77
contribution only.  The scale is the same as in Fig.~1. }}
\end{figure}

The $b$ quark mass can be eliminated in favor of the $B$ meson mass by a change
of variables to
\begin{equation}
  x_B = 2E_\gamma / m_B \,.
\end{equation}
Using $m_b = m_B - \bar\Lambda + (\lambda_1+3\lambda_2)/(2m_b) + \ldots$, 
the photon spectrum becomes
\begin{equation}
\frac{d\Gamma}{dx_B} = 
  \bigg(1 + {\bar\Lambda\over m_B} + \ldots \bigg)\, 
  \frac{d\Gamma}{dx_b} \bigg|_{x_b = x_B(1 + \bar\Lambda/m_B + \ldots)} \,.
\end{equation}
For $x_B$ within a region of order $\Lambda_{\rm QCD}/m_B$ of unity (its
maximal value) nonperturbative effects are very important.  However, for
integrals of $x_B$ over a large enough range these nonperturbative effects are
small. 

An important integral of this type is
\begin{equation}\label{moment1}
\overline{(1 - x_B)} \Big|_{x_B > 1 - \delta} = 
  {\displaystyle \int_{1-\delta}^1 dx_B\, (1-x_B)\, \frac{d\Gamma}{dx_B} \over
   \displaystyle \int_{1-\delta}^1 dx_B\, \frac{d\Gamma}{dx_B} } \,.
\end{equation}
The parameter $\delta=1-2E_\gamma^{\rm min}/m_B$ has to satisfy 
$\delta > \Lambda_{\rm QCD}/m_B$; otherwise nonperturbative effects are not
under control.  It is straightforward to show that
\begin{equation}\label{beauty}
\overline{(1 - x_B)} \Big|_{x_B > 1-\delta} =
  {\bar\Lambda\over m_B} + \bigg(1-{\bar\Lambda\over m_B}\bigg)\,
  \langle 1 - x_b \rangle \Big|_{x_b > 1-\delta}
 - {\bar\Lambda\over m_B}\, \delta(1-\delta)\, \frac1{\Gamma_0}\,
  \frac{d\Gamma}{dx_b} \bigg|_{x_b = 1-\delta} + \ldots \,,
\end{equation}
where 
\begin{equation}\label{momentq}
\langle 1 - x_b \rangle \Big|_{x_b > 1 - \delta} = \int_{1-\delta}^1 dx_b\, 
  (1-x_b)\, \frac1{\Gamma_0}\, \frac{d\Gamma}{dx_b} \,.
\end{equation}
Note that all terms but the first one in Eq.~(\ref{beauty}) have perturbative
expansions which begin at order $\alpha_s$.  The ellipses denote contributions
of order $(\Lambda_{\rm QCD}/m_B)^3$, $\alpha_s(\Lambda_{\rm QCD}/m_B)^2$, and
$\alpha_s^2$ terms not enhanced by $\beta_0$, but it does not contain
contributions of order $(\Lambda_{\rm QCD}/m_B)^2$ or additional
terms\footnote{There are actually additional contributions formally of order
$\alpha_s(\Lambda_{\rm QCD}/m_B)$ coming from the expansion of $m_c/m_b$ in the
22 and 27 terms.  Although the 27 term is very sensitive to the value of
$m_c/m_b$, this $\bar\Lambda$-dependence is negligible for $\overline{(1-x_B)}
|_{x_B>1-\delta}$.} of order $\alpha_s(\Lambda_{\rm QCD}/m_B)$.  Terms in the
operator product expansion proportional to $\lambda_{1,2}/m_b^2$ enter
precisely in the form so that they are absorbed in $m_B$ in
Eq.~(\ref{beauty})~\cite{AZ}.  There are also nonperturbative corrections
suppressed by $(\Lambda_{\rm QCD}/m_c)^2$ instead of $(\Lambda_{\rm
QCD}/m_b)^2$~\cite{1/mc2}.  These do not contribute to Eq.~(\ref{beauty}).

\begin{figure}
\centerline{\epsfysize=7cm\epsffile{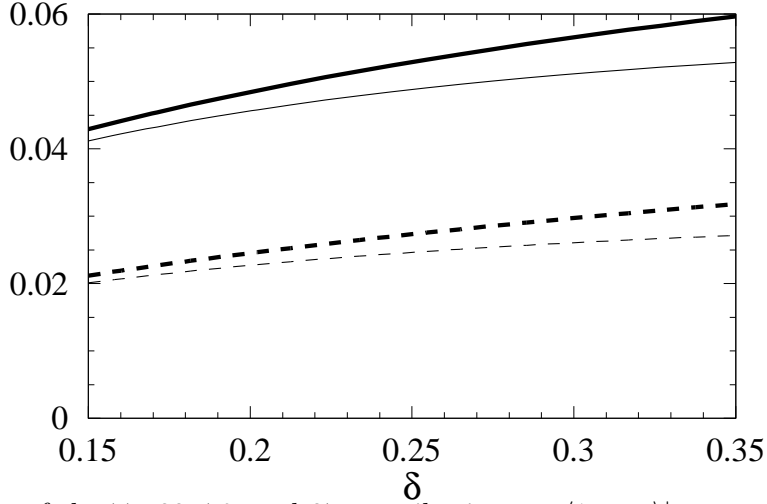}}
\tighten{\caption{The sum of the 77, 22, 78, and 27 contributions to 
$\langle 1-x_b \rangle |_{x_b>1-\delta}$ at order $\alpha_s$ (thick dashed 
curve) and $\alpha_s^2\beta_0$ (thick solid curve).  
The thin curves show the 77 contribution only.}}
\end{figure}

Using our results, $\langle 1-x_b \rangle |_{x_b>1-\delta}$ in
Eq.~(\ref{momentq}) is known to order $\alpha_s^2\beta_0$.  Writing
\begin{equation}
\langle 1 - x_b \rangle \Big|_{x_b > 1-\delta} = 
  B_0(\delta) + \frac{\alpha_s(m_b)}{\pi}\, B_1(\delta)
  + \bigg(\frac{\alpha_s(m_b)}{\pi}\bigg)^2 \beta_0\, B_2(\delta) + \ldots \,,
\end{equation}
$B_p$ have decompositions analogous to Eq.~(\ref{Apdef}),
\begin{equation}
B_p(\delta) = \sum_{i\leq j} b_p^{ij} (\delta) 
  \Bigg[ \frac{C_i(m_b) C_j(m_b)}{C_7^2(m_b)} \Bigg] \,.
\end{equation}
Neglecting $B_0(\delta)$, Eqs.~(\ref{a177}) and (\ref{a277}) yield for the 
dominant 77 contribution
\begin{eqnarray}
b_1^{77} (\delta) &=& \frac{\delta}{54} \left[- 9\delta^3 + 14\delta^2 
  + 72\delta - 54 + 12 (\delta^2 - 3\delta - 6) \ln\delta \right] , \\
b_2^{77} (\delta) &=& \frac{1}{2592} \bigg[ -369\delta^4 + 116\delta^3 +
  1800\delta^2 - 3852\delta\nonumber \\
&+& 408\pi^2 + 12\delta (9\delta^3 + 34\delta^2 - 102\delta + 66) \ln\delta
  \nonumber \\
&-& 216\delta (\delta^2 -3\delta -6) \ln^2 \delta - 144 (\delta^3 - 3\delta^2
  - 6\delta + 17) L_2 (1-\delta)\bigg] \,. \label{b277}
\end{eqnarray}
Our prediction for $\langle 1-x_b \rangle |_{x_b>1-\delta}$ is shown in Fig.~3
as a function of $\delta$, both at order $\alpha_s$ and $\alpha_s^2\beta_0$. 
The bad behavior of the perturbation expansion would improve somewhat by
evaluating the strong coupling at a smaller scale than $m_b$, such as
$m_b\sqrt\delta$, the maximal available invariant mass of the hadronic final
state.  This bad behavior may also be related to the renormalon
ambiguity~\cite{renormalon} in $\bar\Lambda$.

A determination of $\bar\Lambda$ is straightforward using Eq.~(\ref{beauty}). 
The left hand side is directly measurable, while $\langle 1-x_b \rangle
|_{x_b>1-\delta}$ and $(1/\Gamma_0)d\Gamma/dx_b|_{x_b>1-\delta}$ in the second
and third terms on the right hand side can be read off from Figs.~3 and 2,
respectively.  Using the CLEO data in the region $E_\gamma >
2.1\,$GeV~\cite{CLEO}, we obtain the central values
$\bar\Lambda_{\alpha_s^2\beta_0} \simeq 270\,$MeV and $\bar\Lambda_{\alpha_s}
\simeq 390\,$MeV.  We have indicated the order kept in the perturbation
expansion to determine $\bar\Lambda$, since a value of $\bar\Lambda$ extracted
from data can only be used consistently in predictions valid to the same order
in $\alpha_s$.  These values are consistent with the ones obtained from a fit
to the $\bar B\to X_c\,\ell\,\bar\nu$ lepton spectrum~\cite{gremmetal}, and
from the CLEO fit~\cite{CLEO2} to the $\bar B\to X_c\,\ell\,\bar\nu$ hadron
mass distribution~\cite{FLSmass}.

At the present time this extraction of $\bar\Lambda$ has large uncertainties. 
The potentially most serious one is from both nonperturbative and perturbative
terms that are singular as $x\to 1$ and sum into a shape function that modifies
the spectrum near the endpoint.  A model independent determination of these
effects is not available at the present time, however, it may be possible to
address this issue using lattice QCD~\cite{latticeshape}.  For sufficiently
large $\delta$ these effects are not important.  It has been estimated that
they may be significant even if the cut on the photon energy is lowered to
around $E_\gamma = 2\,$GeV~\cite{KaNe,Bauer}, but this is based on
phenomenological models.  We have implicitly neglected these effects throughout
our analysis.  The validity of this can be tested experimentally by checking
whether the value of $\bar\Lambda$ extracted from Eq.~(\ref{beauty}) is
independent of $\delta$ in some range.  This would also improve our confidence
that the total decay rate in the region $x_B>1-\delta$ can be predicted in
perturbative QCD without model dependence.

The value of $\bar\Lambda$ at order $\alpha_s$ has a sizable scale dependence:
lowering the scale such that $\alpha_s$ changes from 0.22 to 0.3 reduces the
value of $\bar\Lambda_{\alpha_s}$ by about $40\,$MeV.  At order
$\alpha_s^2\beta_0$ this scale dependence is much smaller.  Uncertainties due
to the unknown order $(\Lambda_{\rm QCD}/m_B)^3$ terms in the OPE~\cite{Bauer}
are largely uncorrelated to those in the analyses of the lepton energy or
hadron mass spectra in $\bar B\to X_c\,\ell\,\bar\nu$~\cite{GK}.  The effect of
the boost from the $B$ rest frame into the $\Upsilon(4S)$ is small for
$\overline{(1-x_B)} |_{x_B>1-\delta}$~\cite{KaNe}.

\begin{figure}
\centerline{\epsfysize=7cm\epsffile{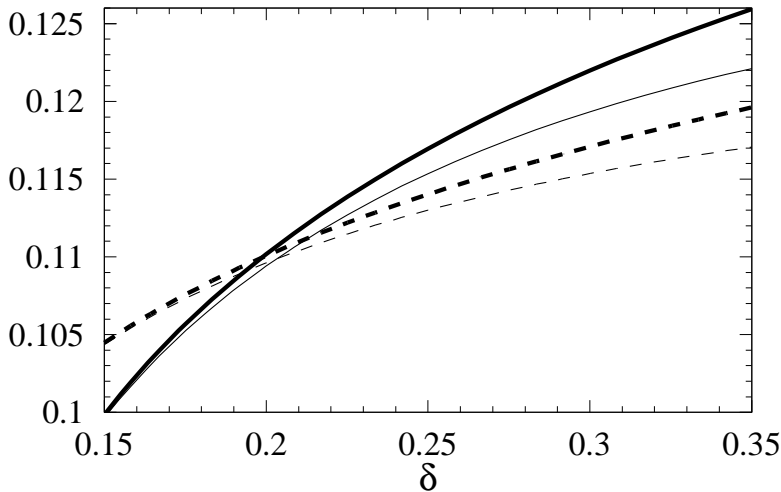}}
\tighten{\caption{Prediction for $\overline{(1 - x_B)} |_{x_B > 1-\delta}$ 
in the upsilon expansion at order $\epsilon$ (thick dashed curve) and 
$(\epsilon^2)_{\rm BLM}$ (thick solid curve).
The thin curves show the 77 contribution only.}}
\end{figure}

The upsilon expansion~\cite{upsexp} yields parameter free predictions for
$\overline{(1 - x_B)} |_{x_B > 1-\delta}$ in terms of the $\Upsilon(1S)$ meson
mass.  The analog of Eq.~(\ref{beauty}) is
\begin{equation}\label{upsbeauty}
\overline{(1 - x_B)} \Big|_{x_B > 1-\delta} = 1 -
  {m_\Upsilon\over 2m_B} \left[ 1 + 0.011\epsilon + 0.019(\epsilon^2)_{\rm BLM}
  - \langle 1-x_b \rangle \Big|_{x_b > (2m_B/m_\Upsilon)(1-\delta)} \right] ,
\end{equation}
where $\epsilon\equiv1$ denotes the order in the upsilon expansion.  For
$E_\gamma>2.1\,$GeV this relation gives 0.111, whereas the central value from
the CLEO data is around 0.093.\footnote{It is interesting to note that
including the CLEO data point in the $1.9\,{\rm GeV} < E_\gamma < 2.1\,$GeV
bin, the experimental central value of $\overline{(1 - x_B)}$ over the region
$E_\gamma > 1.9\,$GeV is 0.117, whereas the upsilon expansion predicts 0.120.}
In Fig.~4 we plot the prediction for $\overline{(1 - x_B)} |_{x_B > 1-\delta}$
as a function of $\delta$, both at order $\epsilon$ and $(\epsilon^2)_{\rm
BLM}$.  The perturbation expansion is much better behaved than the one shown in
Fig.~3.  The most important uncertainty in this approach is the size of
nonperturbative contributions to the $\Upsilon(1S)$ mass other than those which
can be absorbed into the $b$ quark mass.  These have been neglected in
Eq.~(\ref{upsbeauty}).  If the nonperturbative contribution to the
$\Upsilon(1S)$ mass, $\Delta_\Upsilon$, were known, it could be included by
replacing $m_\Upsilon$ by $m_\Upsilon-\Delta_\Upsilon$. For example,
$\Delta_\Upsilon = +300\,$MeV increases $\overline{(1 - x_B)}$ by 21\%, so
measuring $\overline{(1 - x_B)}$ with such accuracy will have important
implications for the physics of quarkonia as well as for $B$ physics.

The variance of the photon energy distribution can be used to determine
$\lambda_1$~\cite{AZ,Bauer}.  The analog of Eq.~(\ref{beauty}) in this case is
\begin{eqnarray}\label{beauty2}
\overline{(1 - x_B)^2} \Big|_{x_B > 1-\delta} - 
  \left[ \overline{(1 - x_B)} \Big|_{x_B > 1-\delta} \right]^2 =
  -{\lambda_1\over 3m_B^2} &+& {\beta^2\over3} +
  \bigg(1-{2\bar\Lambda\over m_B}\bigg)\,
  \langle (1 - x_b)^2 \rangle \Big|_{x_b > 1-\delta} \nonumber\\*
&-& {\bar\Lambda\over m_B}\, \delta^2(1-\delta)\, \frac1{\Gamma_0}\,
  \frac{d\Gamma}{dx_b} \bigg|_{x_b = 1-\delta} + \ldots \,,
\end{eqnarray}
where $\beta\simeq0.064$ is the magnitude of the velocity of the $B$ meson in
the $\Upsilon(4S)$ rest frame, and only the leading $\beta$-dependence has been
kept.  The ellipses denote terms of order $(\Lambda_{\rm QCD}/m_B)^3$,
$\alpha_s(\Lambda_{\rm QCD}/m_B)^2$, and $\alpha_s^2$ terms not enhanced by
$\beta_0$.  Our prediction for $\langle (1-x_b)^2 \rangle |_{x_b>1-\delta}$ is
shown in Fig.~5.  Note that unlike the case of $\overline{(1 - x_B)} |_{x_B >
1-\delta}$, the effect of the boost is very important in Eq.~(\ref{beauty2}). 
Using the CLEO data in the region $E_\gamma > 2.1\,$GeV, we obtain the central
value $\lambda_1 \simeq -0.1\,{\rm GeV}^2$, with large experimental errors. 
The uncertainty in this value of $\lambda_1$ due to $\bar\Lambda$ is small. 
Nonperturbative effects from the cut on $E_\gamma$~\cite{Bauer}, and the
unknown higher order contributions to Eq.~(\ref{beauty2}) are expected to have
a larger impact on the determination of $\lambda_1$ than the corresponding
effects have on the determination of $\bar\Lambda$ from Eq.~(\ref{beauty}).  

\begin{figure}
\centerline{\epsfysize=7cm\epsffile{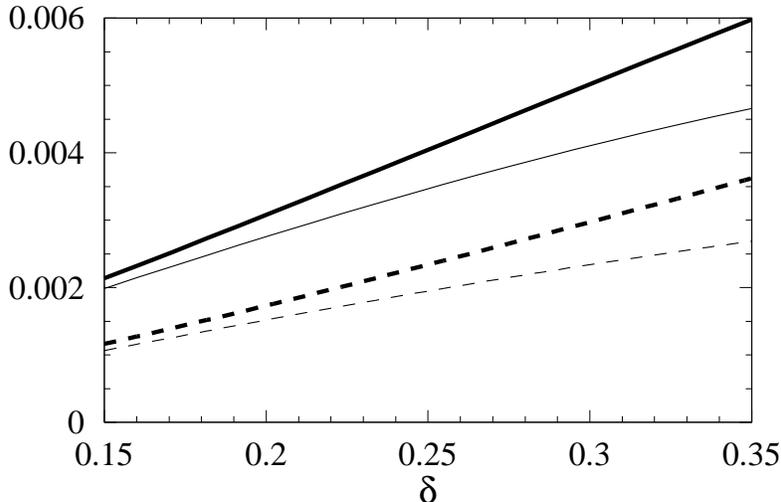}}
\tighten{\caption{The sum of the 77, 22, 78, and 27 contributions to 
$\langle (1-x_b)^2 \rangle |_{x_b>1-\delta}$ at order $\alpha_s$ (thick dashed 
curve) and $\alpha_s^2\beta_0$ (thick solid curve).  
The thin curves show the 77 contribution only.}}
\end{figure}

In summary, we calculated order $\alpha_s^2\beta_0$ corrections to the shape of
the photon energy spectrum in weak radiative $\bar B\to X_s\gamma$ decay.  The
dominant 77 contribution is given by simple analytic formulae in
Eqs.~(\ref{a177}) and (\ref{a277}).  The other terms relevant in the region
$x_b>0.65$ are the 22 and 27 contributions given in Eqs.~(\ref{ai22}) and
(\ref{ai27}), and the 78 term given in Eqs.~(\ref{a178}) and (\ref{a278}).  The
HQET parameter $\bar\Lambda$ can be extracted from the average $\langle 1 -
2E_\gamma/m_B \rangle$ using Eq.~(\ref{beauty}), and it can also be used to
test whether the nonperturbative contribution to the Upsilon mass is small. 
The CLEO data in the region $E_\gamma > 2.1\,$GeV implies the central values
$\bar\Lambda_{\alpha_s} \simeq 390\,$MeV and $\bar\Lambda_{\alpha_s^2\beta_0}
\simeq 270\,$MeV at order $\alpha_s$ and $\alpha_s^2\beta_0$, respectively. 
Possible contributions to the total decay rate from physics beyond the standard
model are unlikely to affect this determination of $\bar\Lambda$.  In the
future, checking the $\delta$-independence of the extracted value of
$\bar\Lambda$, and comparing the experimental and theoretical shapes of the
photon spectrum for $x_b<0.9$ can provide a check that nonperturbative effects
and backgrounds are under control.  This would also improve our confidence that
the total decay rate in the region $x_B > 1-\delta$ can be predicted model
independently, and used to search for signatures of new physics with better
sensitivity.

\acknowledgements 
We thank Mikolaj Misiak for several useful discussions.  M.L.\ thanks the
Caltech Theory Group for hospitality while part of this work was completed. 
This work was supported in part by the Department of Energy under Grant Nos. 
DE-FG03-92-ER40701 and DOE-FG03-97ER40546.  M.L.\ was supported in part by the
Natural Sciences and Engineering Research Council of Canada and the Sloan
Foundation.  Fermilab is operated by Universities Research Association, Inc.,
under DOE contract DE-AC02-76CH03000.

\end{document}